\documentclass[final,5p,times,number,twocolumn,sort&compress]{elsarticle}
\usepackage{lineno}

\usepackage{graphicx}						%
\usepackage{subcaption}							%
\usepackage[]{hyperref}    				 	
\usepackage[all]{hypcap}						
\usepackage{amsmath}                         	
\usepackage{amssymb}                        	
\usepackage{xkeyval}                         	%
\usepackage{xspace}                          	%
\usepackage[alsoload=hep,
	    alsoload=prefixed,
	    alsoload=abbr]{siunitx}   	       	

\hypersetup{
     unicode=false,          
     pdftoolbar=true,        
     pdfmenubar=true,        
     pdffitwindow=false,     
     pdftitle={A laser-based alignment system (LBAS) for nuclear-physics experiments},    		
     pdfauthor={A.M. Rogers},
     pdfsubject={},          
     pdfnewwindow=true,      
     pdfkeywords={}, 		
     colorlinks=true,        
     linkcolor=black,        
     citecolor=blue,         
     filecolor=magenta,      
     urlcolor=blue           
}

\DeclareGraphicsExtensions{.pdf,.png,.jpg}
\graphicspath{{Figures/},{Images/}}

\newcommand{\affANL}{Physics Division, Argonne National Laboratory, Argonne, IL 60439 USA}

\newcommand{\affNSCL}{National Superconducting Cyclotron Laboratory, Michigan State University, East Lansing, MI 48824, USA}

\newcommand{\affWMU}{Department of Physics, Western Michigan University, Kalamazoo, MI 49008, USA}
\newcommand{\affOak}{Oak Ridge National Laboratory, Oak Ridge, TN 37831, USA}
\newcommand{\affUT}{Department of Physics \& Astronomy, University of Tennessee, Knoxville, Tennessee 37996, USA}
\newcommand{\affCAEN}{LPC Caen, ENSICAEN, Universit\'{e} de Caen, CNRS/IN2P3, Caen, France}
\newcommand{\affLanzhou}{Institute of Modern Physics, CAS, Lanzhou 73000, China}

\begin{document}

\begin{frontmatter}



\title{Tracking rare-isotope beams with microchannel plates}


\author[NSCL,ANL]{A.~M.~Rogers}		
\author[NSCL]{A.~Sanetullaev}		
\author[NSCL]{W.G.~Lynch}			
\author[NSCL]{M.B.~Tsang}			
\author[NSCL]{J.~Lee}				
\author[NSCL]{D.~Bazin}				
\author[NSCL]{D.Coupland}			
\author[NSCL]{V.~Henzl}				
\author[NSCL]{D.~Henzlova}			
\author[NSCL]{M.~Kilburn}				
\author[NSCL]{M.~S.~Wallace}			
\author[NSCL]{M.~Youngs}				
\author[CAEN]{F.~Delaunay}			
\author[WMU]{M.~Famiano}				
\author[Oak]{D.~Shapira}				
\author[UT]{K.~L.~Jones}				
\author[UT]{K.~T.~Schmitt}			
\author[Lanzhou]{Z.~Y.~Sun}			

\address[NSCL]{\affNSCL}
\address[ANL]{\affANL}
\address[CAEN]{\affCAEN]}
\address[WMU]{\affWMU}
\address[Oak]{\affOak}
\address[UT]{\affUT}
\address[Lanzhou]{\affLanzhou}

\begin{abstract}
A system of two microchannel-plate detectors has been successfully implemented for tracking projectile-fragmentation beams.
The detectors provide interaction positions, angles, and arrival times of ions at the reaction target.
The current design is an adaptation of an assembly used for low-energy beams ($\sim$1.4~MeV/nucleon).
In order to improve resolution in tracking high-energy heavy-ion beams, the magnetic field strength between the secondary-electron accelerating foil and the microchannel plate had to be increased substantially.
Results from an experiment using a 37-MeV/nucleon ${}^{56}$Ni beam show that the tracking system can achieve sub-nanosecond timing resolution and a position resolution of $\sim$1~mm for beam intensities up to $5\times10^{5}$~pps. 
\end{abstract}


\end{frontmatter}



\section{Introduction}
\label{sec:Intro}
Many modern nuclear-physics facilities utilize secondary rare-isotope beams to investigate the properties of neutron-rich and neutron-deficient nuclei~\cite{Dufour:1986iq,Kubo:1992jq,munzenberg,Sherrill:1997hu,NationalResearchCouncilUSRareIsotopeScience:2007wq,Lecesne:2008ke}.
Studies that employ radioactive-ion beams are crucial for understanding topics ranging from astrophysics to quantum many-body systems such as the atomic nucleus~\cite{NationalResearchCouncilUSRareIsotopeScience:2007wq}.
Facilities that use projectile fragmentation or fission, produce secondary beams composed of a broad range of isotopes.
Specific isotopes in these beams are selected using fragment separators that utilize combinations of electric fields, magnetic fields, and degraders~\cite{Geissel:1992ji,1995ARNPS..45..163G,2003NIMPB.204...90M}.
The resulting beams, however, typically have a large emittance arising from the broad momentum distributions of the secondary-beam particles produced in fragmentation or through the fission process as well as scattering and straggling introduced by the degraders.
In experiments requiring high-resolution position and energy data, event-by-event measurements of the momenta of secondary-beam particles are needed to compensate for this intrinsically large beam emittance~\cite{1997PhRvC..55..562Z,2010PhRvL.104k2701L,2011PhRvC..83a4606L,2011PhRvL.106y2503R,Sanetullaev:vv}.

Several related tracking-detector systems have been developed previously~\cite{OttiniHustache:1999gg,Shapira2000409,Bougamont:2004gc,Hashimoto:2006ek,Drouart:2007hv}.
Most of them are rate limited on the order of $\sim$\numrange[retain-unity-mantissa = false, range-phrase = --]{1e3}{1e4} particles per second (pps).
In this paper we report on the performance of position-sensitive microchannel-plate (MCP) detectors that can handle rates as high as \SI{5e5}{pps}.
Since our MCP system detects and amplifies secondary electrons emitted by fast beam ions passing through a thin foil, problems associated with beam-induced background reactions, energy losses, and the effects of multiple scattering are minimized~\cite{Wiza:1979ti}.
While the efficiencies involved in detecting light particles using thin foils are low, the efficiencies of the system for isotopes with $Z>20$ are found to approach unity~\cite{Shapira2000409}.

This paper is organized as follows.
In Section 2 we describe the MCP system used in our measurements.
In Sections 3 and 4 we demonstrate its performance both with $\alpha$ particles from a radioactive source and heavy-ion beams, respectively.
In Section 5 we describe the implementation of the microchannel-plate tracking detectors in fast-beam experiments and show how they improve the angular and energy resolution of transfer-reaction data.
Finally, we summarize our findings in Section 6.

\section{Position sensitive microchannel-plate detectors}
\label{sec:PosMCPs}
\subsection{Microchannel plates}
\label{sec:MCPs}
Microchannel plates are compact arrays of single-channel electron multipliers~\cite{Wiza:1979ti,1966ITNS...13...88A}.
As described in Ref.~\cite{Wiza:1979ti}, these electron multipliers are glass tubes, with individual diameters of about \SI{30}{microns} or less, that have been fused together in a closed packed array and subsequently cut into wafers of the order of a millimeter in length.
The front and back surfaces of these wafers are coated with a conductive surface that allows a uniform bias to be applied across the MCP and consequently between the two ends of each tube or ``channel'' in the MCP.
Electrons that enter a channel and strike its interior surface can produce secondary electrons that are accelerated down the channel and multiplied.
The efficiency for detection depends on the angle and energy of incidence, reaching a peak at energies of about a few hundred eV~\cite{Wiza:1979ti}.
The gain of such a channel depends on the average number of collisions inside the channel and is largely governed by the ratio of its length to its diameter~\cite{1966ITNS...13...88A}.
Progress in fabrication techniques has made it possible to scale this technology to produce channels with diameters as small as \SI{2}{\micron} and length/width ratios of up to \num{80}~\cite{website:photonis}.
Gain within a single channel can be limited by space charge, but gain factors of $>\num[retain-unity-mantissa = false]{1e7}$ can be reached by stacking three or more MCP wafers in a Z-stack configuration~\cite{Firmani:1982uz}.
The large gains of MCPs make them an excellent option for applications requiring single electron counting.
The short length of the individual channels gives them the property of being excellent timing detectors~\cite{Akatsu:2004ia}.
As a result of their narrow channels, MCPs are relatively immune to magnetic field effects at low magnetic fields, however, gain reductions of three to five have been observed in axial magnetic fields of \SI{2}{T}~\cite{Barnyakov:2007gv}.

\begin{figure}
  \centering
  \includegraphics[width=0.450\textwidth]{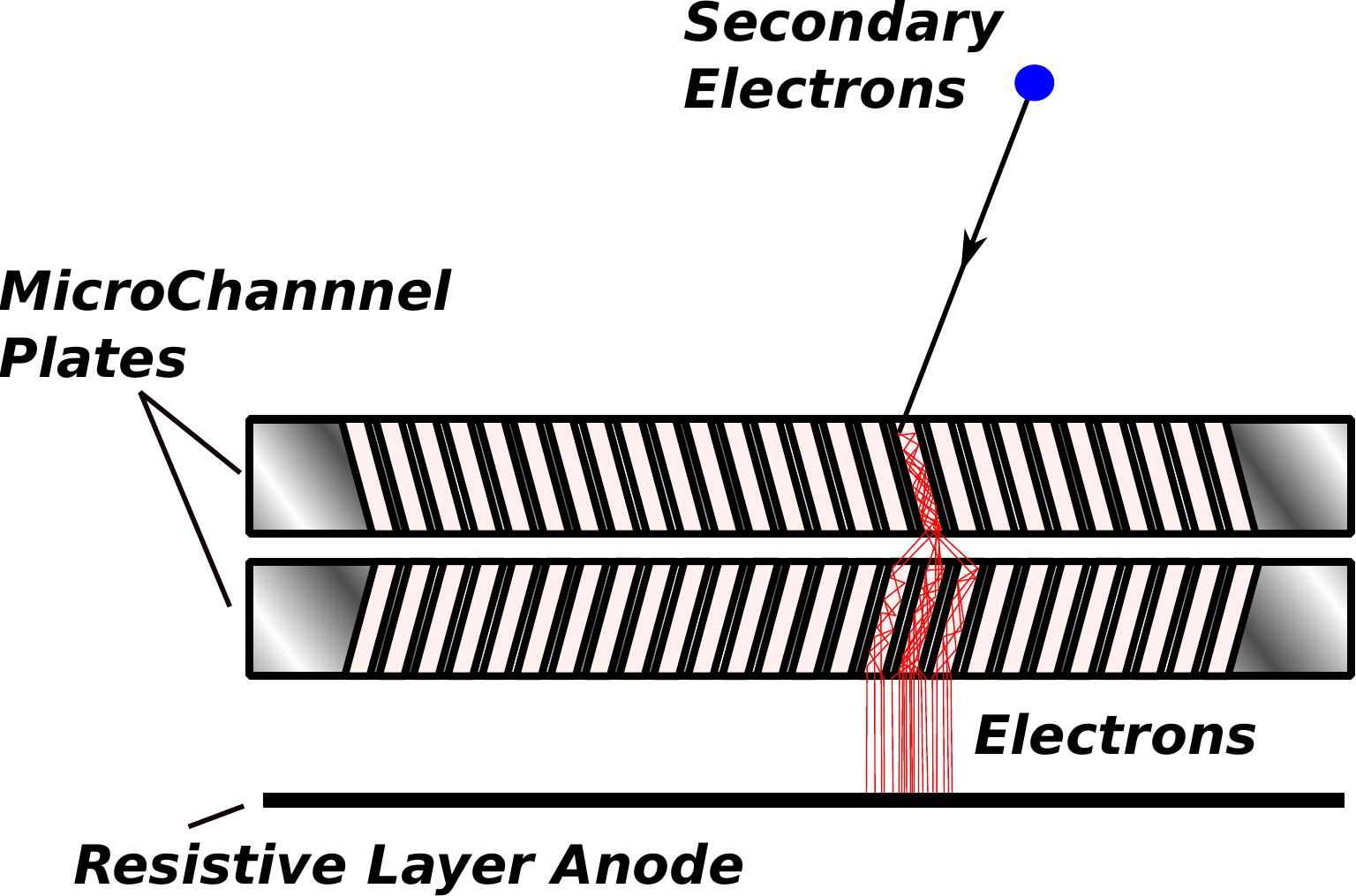}
  \caption{(color online). Signal amplification from secondary-electron emission with two chevron-design microchannel plates.}
  \label{fig:McpAval}
\end{figure}

In our configuration, both large gain and stability is enhanced by stacking two MCPs in a Chevron design as shown in \autoref{fig:McpAval}.
In addition to increasing the gain of the system due to the larger number of multiplication steps, stacking can prevent positive ions that are generated at the output end of the channel from flowing back toward the input end and producing delayed pulses.
Such delayed pulses are minimized by forcing the ions to undergo a significant change in direction at the juncture between the two MCPs~\cite{Wiza:1979ti}.

\subsection{Detector setup}
\label{sec:DetectorSetup}
\begin{figure*}
  \centering
  {\includegraphics[width=0.45\textwidth]{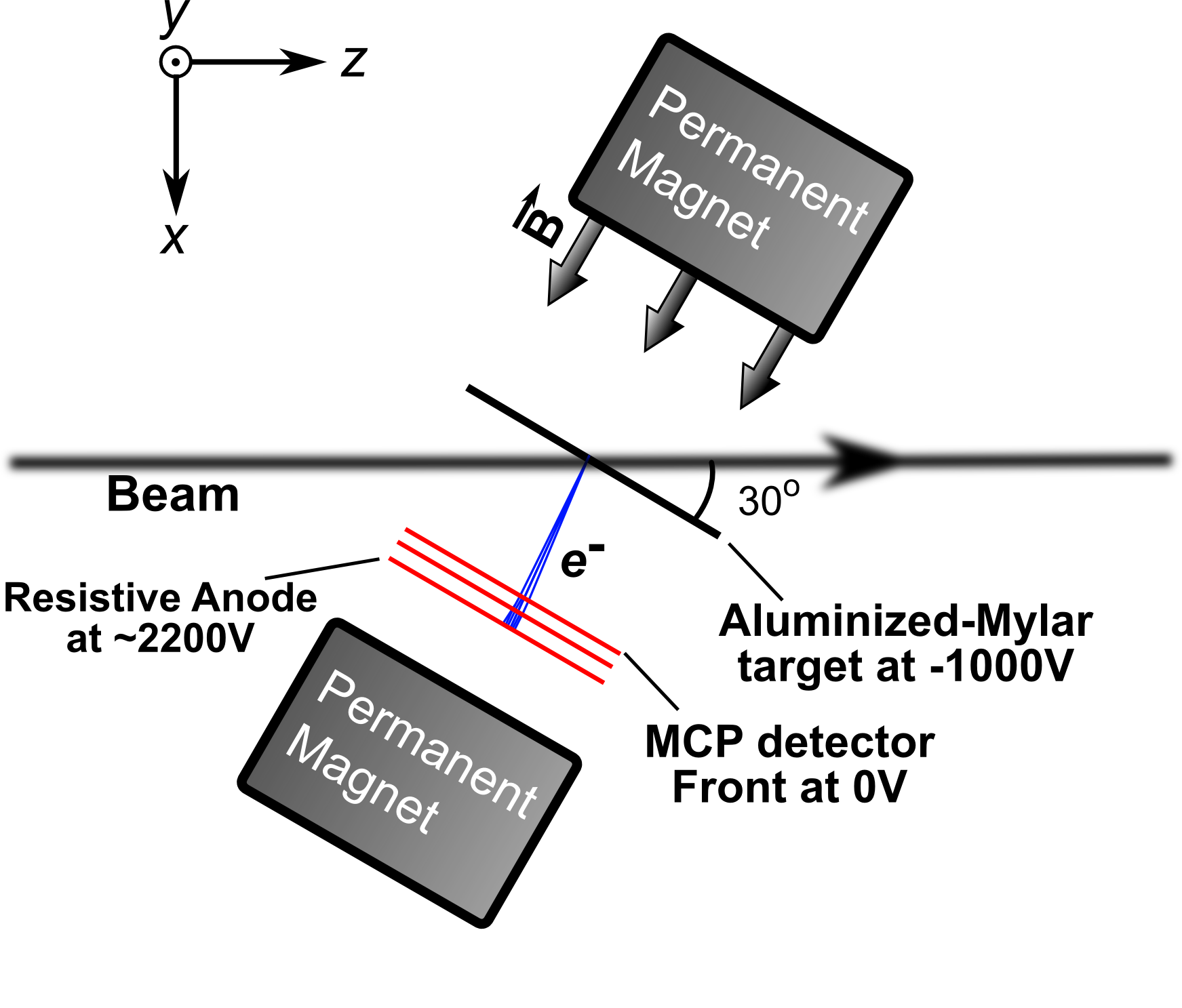}
   \phantomsubcaption\label{a}}
  {\includegraphics[width=0.45\textwidth]{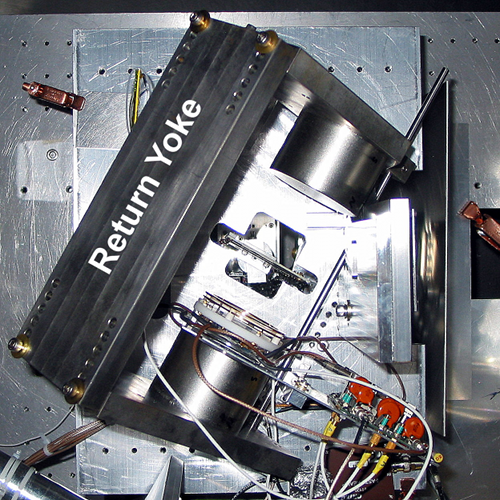}
   \phantomsubcaption\label{b}}
  \caption{(Color online). (Left) Schematic and (right) photographic top view of a MCP tracking detector. The magnetic-flux return yoke used in the current detector setup is visible in the photo.}
  \label{fig:setup}
\end{figure*}

The design of the position-sensitive MCPs used in this work is based on detectors previously described by Shapira~et~al.~\cite{Shapira2000409,Shapira2000396}, which were optimized to detect low-energy ions ($\sim$\SI{1.4}{MeV/nucleon}).
A schematic layout as well as a photograph of a single MCP tracking detector is shown in \autoref{fig:setup}.

When an ion, traveling from left to right in \autoref{fig:setup}, traverses the MCP foil, electrons are scattered from its surface and are multiplied in the MCP.
A thin \num{290}-\si{\ug/cm^2} aluminized Mylar foil is mounted to an insulating fiberglass-composite frame and directly connected by an electrode to a bias-voltage supply.
The entire tracking-detector structure is rotated by \ang{60} to allow the beam to pass through the tracking assembly.
A bias voltage of \SI{-1000}{V} provides an electric field of $\sim\SI{2e4}{V/m}$ that accelerates secondary electrons ejected from the foil towards the MCP whose front surface is at ground potential.
A set of high-voltage divider resistors supply the correct biases to the individual plates where the overall MCP operating voltage is adjusted between \SIrange[retain-explicit-plus]{+2000}{+2200}{V}.
Secondary electrons entering the microchannels are multiplied through an electron avalanche and deposited on the anode behind the MCPs.
Two permanent rare-earth magnets provide a magnetic field that is roughly parallel to the electric field, confining the electrons to tight helical orbits spiraling around the magnetic field lines.
The result is an ``image'' of the beam-interaction point at the foil projected onto the front surface of the MCP via the emitted secondary electrons~\cite{Shapira2000409,Shapira2000396,Culhane19911,Odland1996149}.

We purchased the microchannel plates in a Chevron configuration and assembled with a circular-arc terminated resistive anode, which is position sensitive, from Quantar~ \cite{company:quantar,Lampton:1979wv}.
Each plate is \num{0.46}-\si{mm} thick, \SI{40}{mm} in active diameter, and composed of \num{10}-\si{\um} diameter channels with a spacing of \SI{12}{\um} between centers of adjacent channels.
The channels are biased at an angle of \ang{8} relative to the surface normal.
The anode layer is located \SI{4}{mm} behind the second MCP.
Operation of this device requires a vacuum pressure better than \SI{5e-6}{torr} to prevent discharge of electrons in the microchannel tubes.
Similar setups have been implemented successfully in several nuclear physics experiments~\cite{2011PhRvC..83a4606L,2011PhRvL.106y2503R,Sanetullaev:vv,2011PhRvL.107q2503E}.

We should note that the MCP tracking-detector setup used here for fast heavy-ion beams differs from the previous work of Shapira et al.~\cite{Shapira2000409} in that we do not employ any accelerating grids.
In the work of Shapira et al. accelerating grids were placed at the target foil to create a fast acceleration region for the purpose of rapidly imparting large longitudinal velocities to the electrons to minimize the effect of lateral electron drift on position resolution.
Both the grids at the MCP foil and MCP detector were used to explore the effects on electron transport.
It was found, however, that the addition of a magnetic field would have a more significant effect.
We therefore chose to forgo the complication of accelerating grids in light of this work as well as to remove any excess material from the beam and electron-transport paths.

\subsection{Position determination from resistive charge division}
\label{sec:chargeDiv}
The MCP position signals are encoded by a resistive anode. Capacitively-decoupled position signals are taken from the Upper Left (UL), Lower Left (LL), Upper Right (UR) and Lower Right (LR) corners of the anode.
The amplitude of each anode corner signal is proportional to the number of secondary electrons produced in the microchannel plates and inversely proportional to the resistance between the charge-deposition location on the resistive layer and the corner electrode.
Thus, the amplitude of a corner signal will be greater if the deposition is closer to that corner.
Conversely, it will have a lower amplitude if the signal originates from a point farther away.

\begin{figure}
  \centering
  \includegraphics[width=0.450\textwidth]{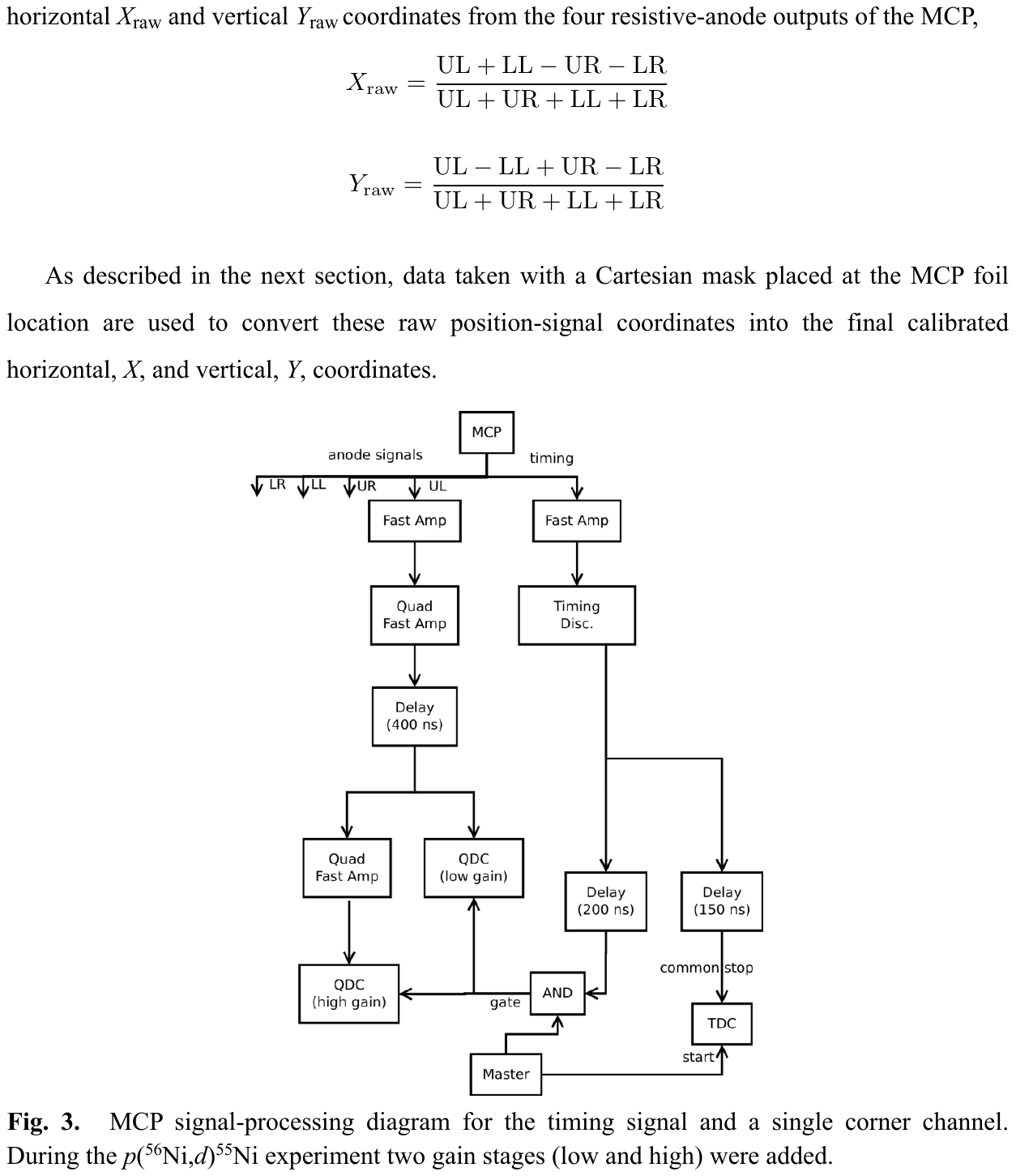}
  \caption{MCP signal-processing diagram for the timing signal and a single corner channel.  During the $p$(${}^{56}$Ni,$d$)${}^{55}$Ni experiment two gain stages (low and high) were implemented.}
  \label{fig:MCPFlow}
\end{figure}

A diagram of the electronics setup used to process the corner signals and the timing signal, T, is shown in \autoref{fig:MCPFlow}.
The capacitively-decoupled timing signal is picked off directly from the biasing electrode of the second microchannel plate and amplified by an Ortec VT120 fast-timing preamplifier which is placed close to the MCP detector inside the scattering chamber.
After the preamplifier, the signal is sent to a Phillips 715 constant-fraction discriminator (CFD) located outside the chamber.
The discriminator output is digitized by a CAEN V1920N Time-to-Digital Converter (TDC) and also used to generate gates for the CAEN N792 Charge-to-Digital Converters (QDC).
A ``master'' event trigger establishes a coincidence between the MCP and events occurring in additional detectors in the setup.

The circuits for the corner signals UL, UR, LL, and LR are identical.
Each corner signal is amplified by a factor of \num{100} in an ORTEC 820 fast amplifier.
In the most recent measurements, two gain stages are employed to accommodate the wide dynamic range of the position signals.
Each position signal from the fast amplifier is split, with a low-gain branch sent directly to a QDC, where it is digitized.
A second high-gain branch is further amplified by a factor of eight using an NSCL fast amplifier~\cite{workOrder:NSCLFA} before being digitized by another QDC.
Signals from both gain stages are matched with a precision pulser and verified using experimental data.
We have found that using two gain stages alleviates problems with position reconstruction near the anode center, where the corner signal amplitudes are the smallest. 

The relative amplitudes of the digitized corner signals provide horizontal- and vertical-position information.
We employ charge division techniques to obtain the raw (uncalibrated) horizontal $X_{\textrm{raw}}$ and vertical $Y_{\textrm{raw}}$ coordinates from the four resistive-anode outputs of the MCP,
\begin{equation}
 X_{\textrm{raw}} = \frac{(\textrm{UL}+\textrm{LL})-(\textrm{UR}-\textrm{LR})}{(\textrm{UL}+\textrm{UR}+\textrm{LL}+\textrm{LR})}
\end{equation}
\begin{equation}
 Y_{\textrm{raw}} = \frac{(\textrm{UL}-\textrm{LL})+(\textrm{UR}-\textrm{LR})}{(\textrm{UL}+\textrm{UR}+\textrm{LL}+\textrm{LR})}
\end{equation}
As described in the next section, data taken with a Cartesian mask placed at the MCP foil location are used to convert these raw position-signal coordinates into the final calibrated horizontal, $X$, and vertical, $Y$, coordinates.

\section{MCP mask calibrations with $\alpha$ particles}
\label{sec:MCPMaskAlpha}
\begin{figure}
  \centering
  \includegraphics[width=0.30\textwidth]{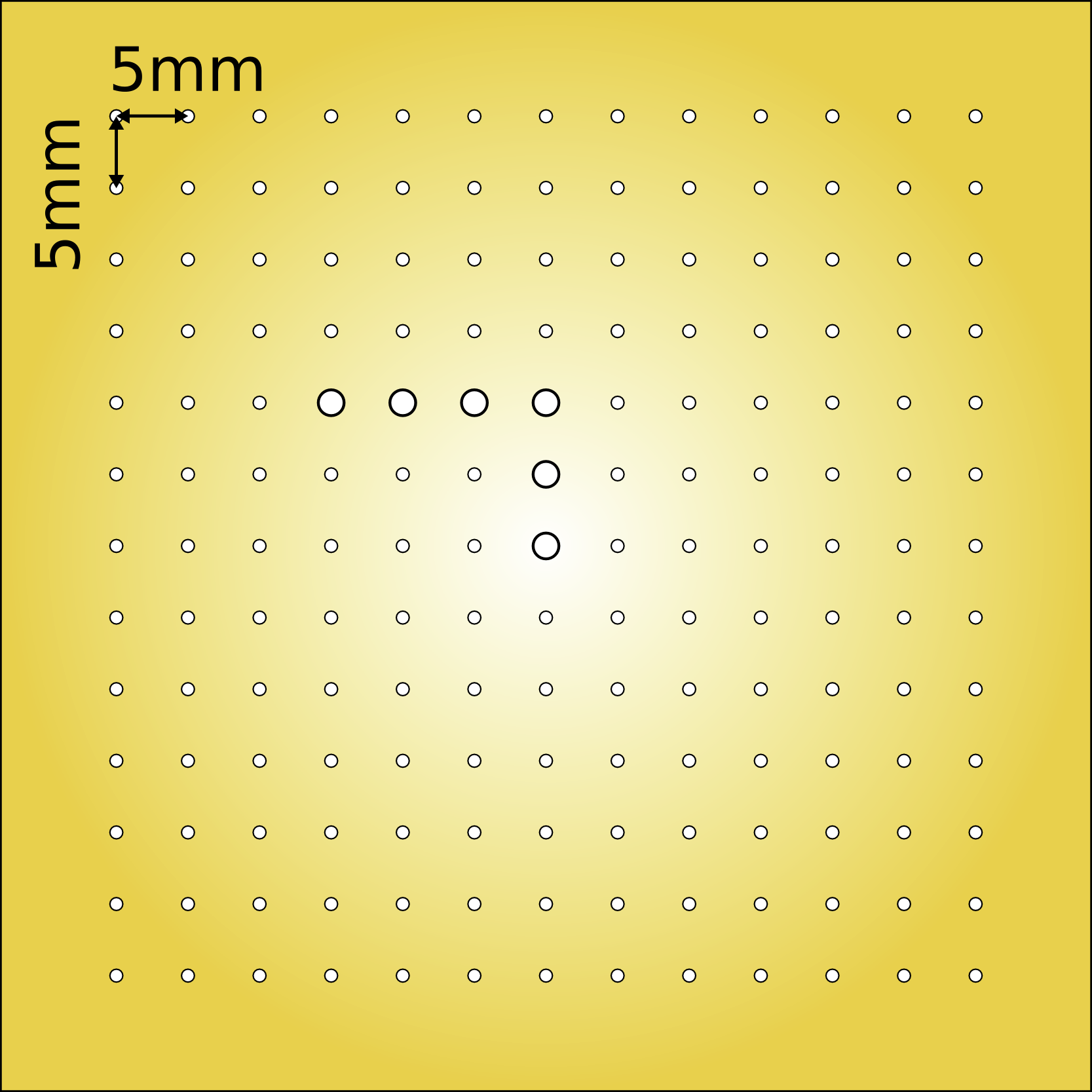}					
  \caption{Brass mask used for position calibration of the MCPs.  The larger holes in the center of the mask are used to determine orientation.}
  \label{fig:Mask}
\end{figure}

A mask made of a \num{0.16}-\si{mm} thick brass plate with a matrix of holes as illustrated in \autoref{fig:Mask}, was placed at the MCP foil position.
The smaller holes in the mask have a diameter of \SI{0.75}{mm}.
The L-shaped pattern on the mask, composed of \num{1.5}-\si{mm} holes, serves to confirm the orientation of the mask.
A \num{290}-\si{\ug/cm^2} aluminized Mylar foil is attached to the mask.
The foil-covered mask is mounted on an insulating fiberglass-composite frame and biased to \SI{-1000}{V}, which again provides the electric field that accelerates the secondary electrons ejected from the foil toward the MCP.

\begin{figure}[t]
  \centering
  {\includegraphics[width=0.40\textwidth]{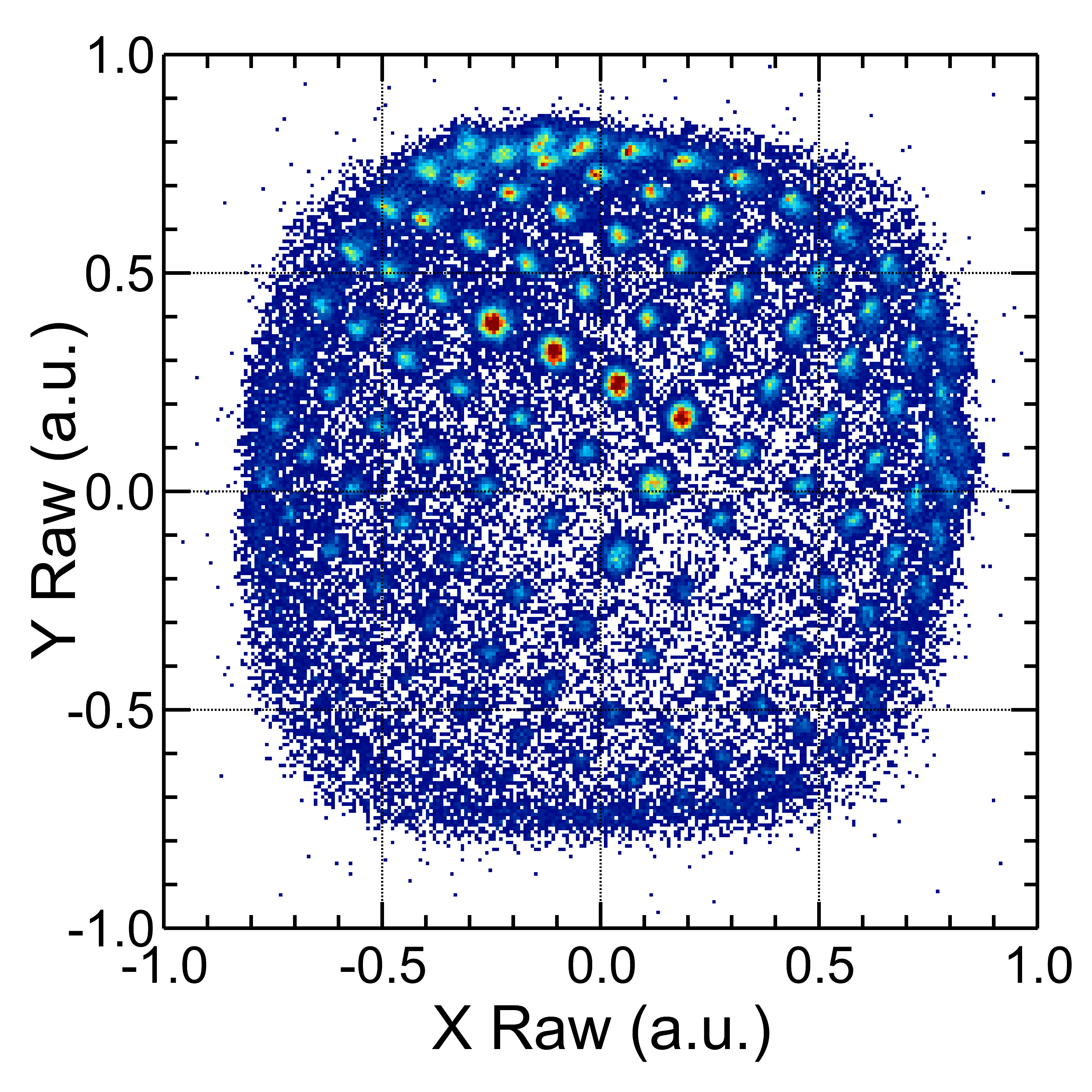}
   \phantomsubcaption\label{a}}
  {\includegraphics[width=0.40\textwidth]{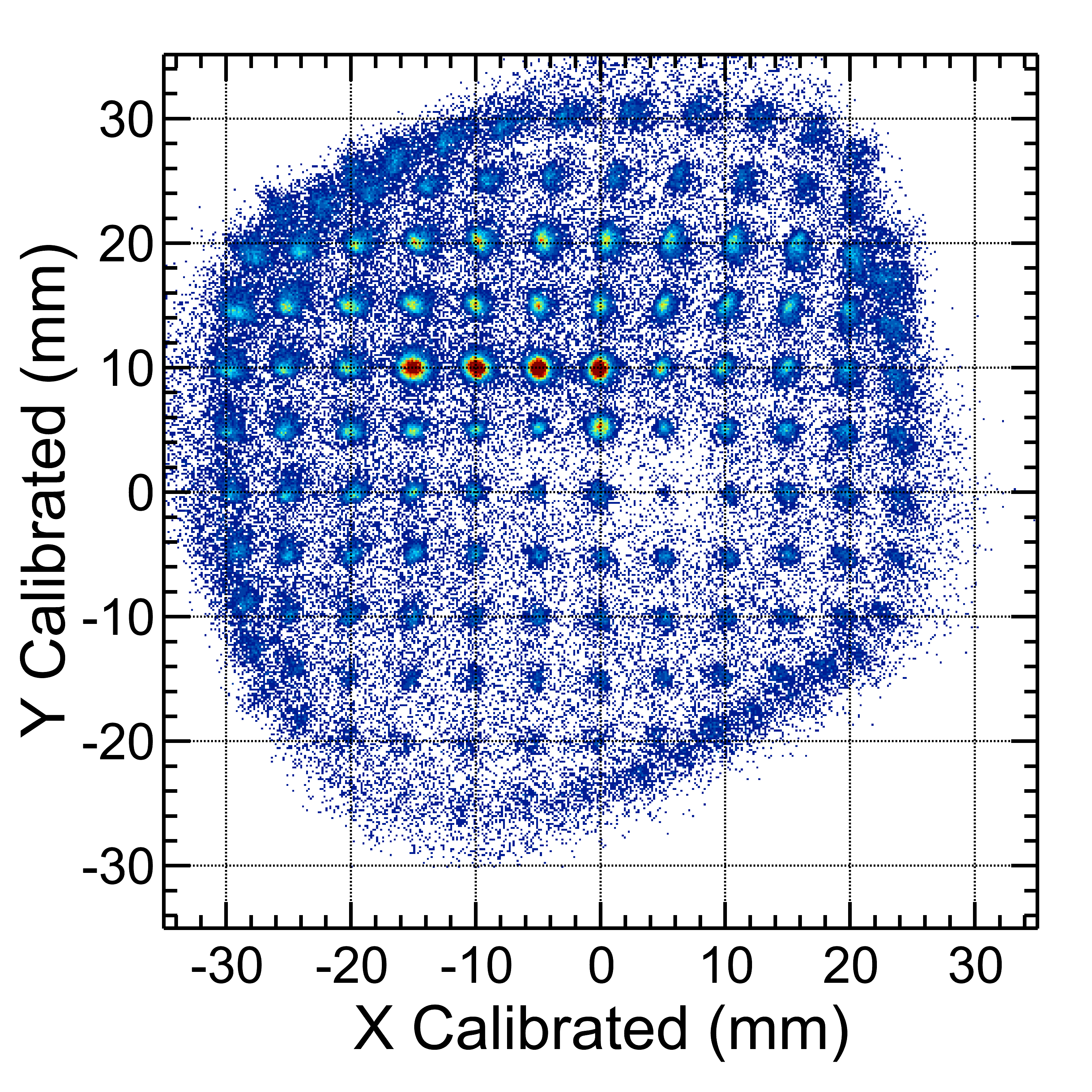}
   \phantomsubcaption\label{b}}
  \caption{(Color online).  (Top) Raw and (bottom) calibrated MCP $\alpha$-source position spectra with magnetic fields of $B_{\textrm{MCP}} = \SI{0.05}{T}$ and $B_{\textrm{foil}}Ê=Ê\SI{0.03}{T}$.}
  \label{fig:AlphaCalib}
\end{figure}

In our first test, a ${}^{228}$Th source was used to develop the position calibration procedure and to determine the position resolution.
This source emits $\alpha$ particles with energies ranging from approximately \SI{5.4}{MeV} to \SI{8.78}{MeV}.
At these energies $\alpha$ particles are stopped in the mask but pass through the aluminized Mylar foil covering the holes with an energy loss of $\sim$\num{200}-\SI{300}{keV}.
Coincidence measurements with a plastic scintillator paddle placed on the opposite side of the mask from the ${}^{228}$Th source allowed us to identify the $\alpha$ particles passing through the holes in the mask.
The uncalibrated and calibrated $\alpha$-source mask data, shown in \autoref{fig:AlphaCalib}, was taken with an MCP detector setup consisting of two permanent magnets without a magnetic-flux return yoke.
These magnets create a field of $B_{\textrm{MCP}} = \SI{0.05}{T}$ at the surface of the MCP and $B_{\textrm{foil}} = \SI{0.03}{T}$ at the MCP foil, as measured with a Gauss meter.
	The magnetic field is smaller at the foil location than it is at the anode position of the MCP.
As the electrons follow tight spirals along the magnetic field lines, the image of the mask is compressed when it is projected onto the MCP.
As a result the mask, which contains holes covering an area of $\SI{60}{mm}\times\SI{60}{mm}$, can be imaged on an anode with an active-area diameter of \SI{40}{mm}.

The procedure we followed to transform the uncalibrated position spectra, $X_{\textrm{raw}}$ and $Y_{\textrm{raw}}$, into calibrated physical Cartesian positions consists of the following steps:
\begin{enumerate}
	\item Pedestals, or channels corresponding to signals with zero amplitude, are initially determined by a special pedestal run and subtracted from the raw data so that the average QDC channel values for UL, UR, LL and LR are zero when no signals are present.
	\item The initial pedestal values are fine-tuned to minimize the dependence of the calculated position of selected holes in the calibration mask on the sum of the corner signals.
	Typically, smaller signals are generated at the center partly due to radiation damage of the central part of the MCP and possibly from loading of the MCP, both of which can be caused by the more intense beam rate experienced by the MCP at its center.
	Thus, the pedestal settings affect the position spectra in the center of the MCP more severely.
	The upper spectrum in \autoref{fig:AlphaCalib} shows the raw position spectrum for an MCP after determining and tuning the best pedestal settings.
	In this spectrum, the L-shaped pattern corresponding to the larger holes in the calibration mask is rotated and the spacing between neighboring holes is not uniform.
	\item A two-dimensional third-order polynomial is used to fit the raw position information to the mask pattern.
	The fit is optimized to correct the uncalibrated positions to match the known physical positions of the holes in the mask as given by,
	\begin{align*}
 X_{\textrm{cal}} = & a_{1} + a_{2}X_{\textrm{raw}} + a_{3}Y_{\textrm{raw}} + a_{4}X^{2}_{\textrm{raw}} + a_{5}Y^{2}_{\textrm{raw}} + a_{6}X^{3}_{\textrm{raw}} \\
  + & a_{7}Y^{3}_{\textrm{raw}} + a_{8}X_{\textrm{raw}}Y_{\textrm{raw}} + a_{9}X^{2}_{\textrm{raw}}Y_{\textrm{raw}} + a_{10}Y^{2}_{\textrm{raw}}X_{\textrm{raw}}
	\end{align*}
	\begin{align*}
 Y_{\textrm{cal}} = & b_{1} + b_{2}Y_{\textrm{raw}} + b_{3}X_{\textrm{raw}} + b_{4}Y^{2}_{\textrm{raw}} + b_{5}X^{2}_{\textrm{raw}} + b_{6}Y^{3}_{\textrm{raw}} \\
 + & b_{7}X^{3}_{\textrm{raw}} + b_{8}Y_{\textrm{raw}}X_{\textrm{raw}} + b_{9}Y^{2}_{\textrm{raw}}X_{\textrm{raw}} + b_{10}X^{2}_{\textrm{raw}}Y_{\textrm{raw}}
	\end{align*}
	where $X_{\textrm{raw}}$, $Y_{\textrm{raw}}$, $X_{\textrm{cal}}$ and $Y_{\textrm{cal}}$ are the raw and calibrated coordinates respectively and $a_{i}$, $b_{i}$ are the fit parameters.
\end{enumerate}

The position spectrum after fitting and calibration is shown in the lower panel of \autoref{fig:AlphaCalib}.
For holes at the center of the mask, the centroids of the peaks are located to within \SI{0.5}{mm} of their correct physical positions.
Individual hole positions cannot be resolved near the edge of the channel plate active area, where the centroids of the peaks can differ by as much as \SI{2.0}{mm}.
Normally, this region and the corner regions lies outside of the area illuminated by most secondary beams.

\section{Performance of microchannel-plate with fast heavy-ion beams}
\label{sec:PerfomMCPHeavyIon}
In this section we discuss the performance of the MCPs with rare-isotope beams used in two experiments conducted at the National Superconducting Cyclotron Laboratory (NSCL) at Michigan State University~\cite{2011PhRvL.106y2503R,Sanetullaev:vv}.
Both experiments involve radioactive beams which contain $\sim\SI{70}{\percent}$ of the beam of interest, while the remaining $\sim\SI{30}{\percent}$ is composed of other isotopes with the same $Q/A$ ratio.
This mixture of beam is termed a ``cocktail beam''.
In the next section, details regarding beam-particle identification will be discussed.
To evaluate the performance of the MCP, however, beam selection is not required since the properties of each beam component are similar, resulting in the emission of nearly identical secondary electrons from the foil.
The position calibrations discussed and spectra shown in this section use data generated by all components of the beams.
Including all beam particles increases our statistics, thereby improving the overall calibration.

The first experiment involved a ${}^{70}$Se cocktail beam containing approximately \SI{67}{\percent} ${}^{70}$Se, \SI{9}{\percent} ${}^{71}$Br, and \SI{24}{\percent} ${}^{69}$As at an energy of \SI{72}{MeV/nucleon} and used the same MCP tracking detectors as those for the $\alpha$-source test described in section 3.
The MCPs were equipped with identical magnets generating field strengths of $B_{\textrm{MCP}} =Ê\SI{0.05}{T}$ and $B_{\textrm{foil}}Ê=Ê\SI{0.03}{T}$.

The second experiment used a \num{37}-\si{MeV/nucleon} ${}^{56}$Ni cocktail beam containing about \SI{71}{\percent} ${}^{56}$Ni, \SI{24}{\percent} ${}^{55}$Co, \SI{2}{\percent} ${}^{54}$Fe, and \SI{2}{\percent} ${}^{53}$Mn.
Stronger neodymium iron boron (NdFeB) magnets \SI{3}{''} in diameter and \SI{2}{''} in length (Magnet Sales and Manufacturing Inc., Part No. 35NERR192) were mounted on magnetic-flux return yokes resulting in fields of $B_{\textrm{MCP}} =Ê\SI{0.21}{T}$ at the MCP and $B_{\textrm{foil}}Ê=Ê\SI{0.13}{T}$ at the target foil~\cite{company:msm}.

\subsection{Position calibration with ${}^{70}$Se cocktail beam at \SI{72}{MeV/nucleon}}
\label{sec:PosCal70Se}
Since the \num{0.16}-\si{mm} thick brass mask in \autoref{fig:Mask} does not stop the beam, a scintillator is inadequate to distinguish particles passing through the holes in the mask from those passing through the brass.
Instead, the NSCL S800 spectrograph Ñ a high-resolution, large-acceptance mass spectrometer composed of superconducting magnets Ñ located downstream of the target directly after the scattering chamber was used~\cite{bazin2003ss}.
The magnetic rigidity of the S800 was set to select only beam particles that pass through the mask holes, allowing them to be detected at the S800 focal plane.
Beam particles that pass through the brass portion of the mask lose sufficient energy so that their trajectories are deflected off the focal-plane detector of the S800 spectrometer.

\begin{figure}[t]
  \centering
  {\includegraphics[width=0.40\textwidth]{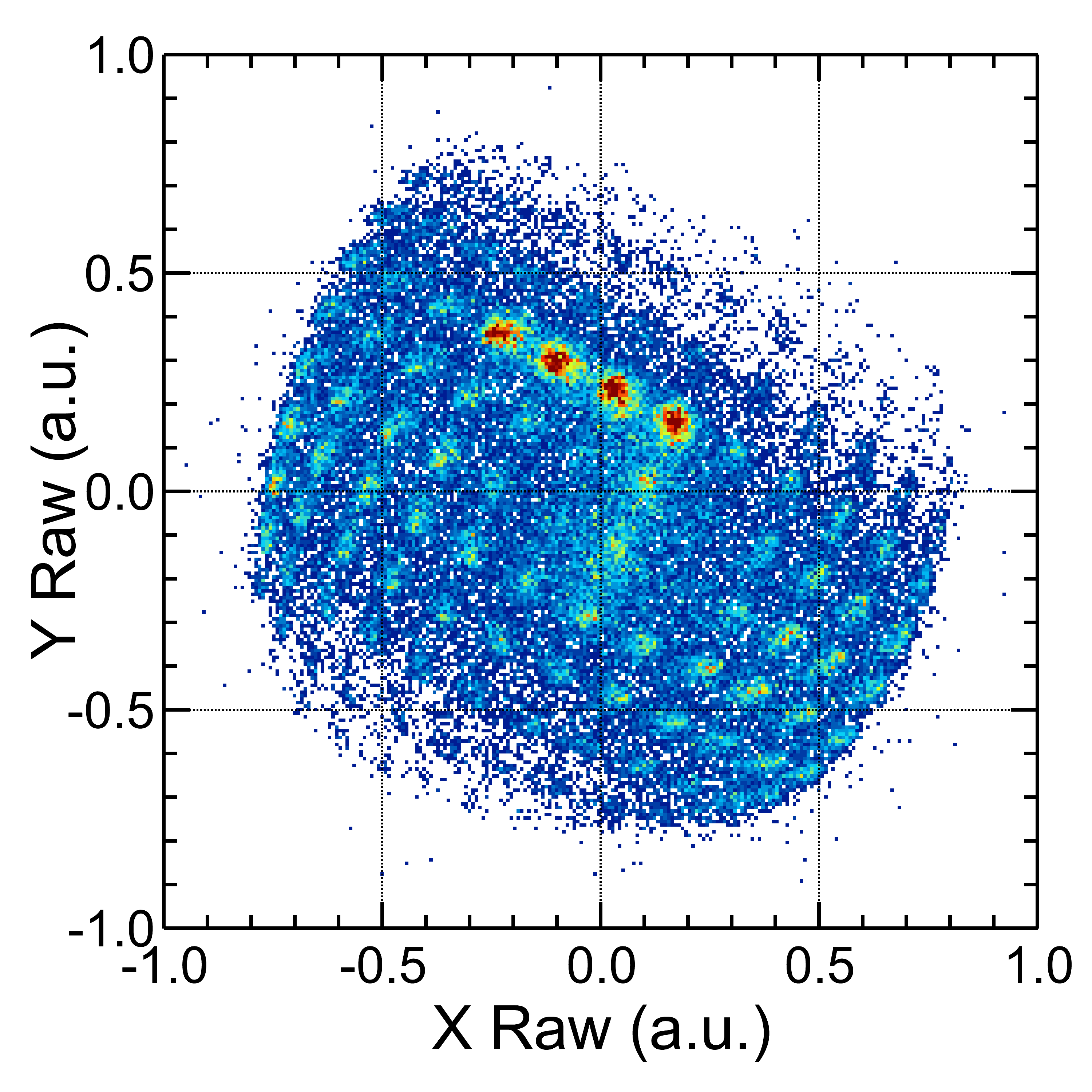}
   \phantomsubcaption\label{a}}
  {\includegraphics[width=0.40\textwidth]{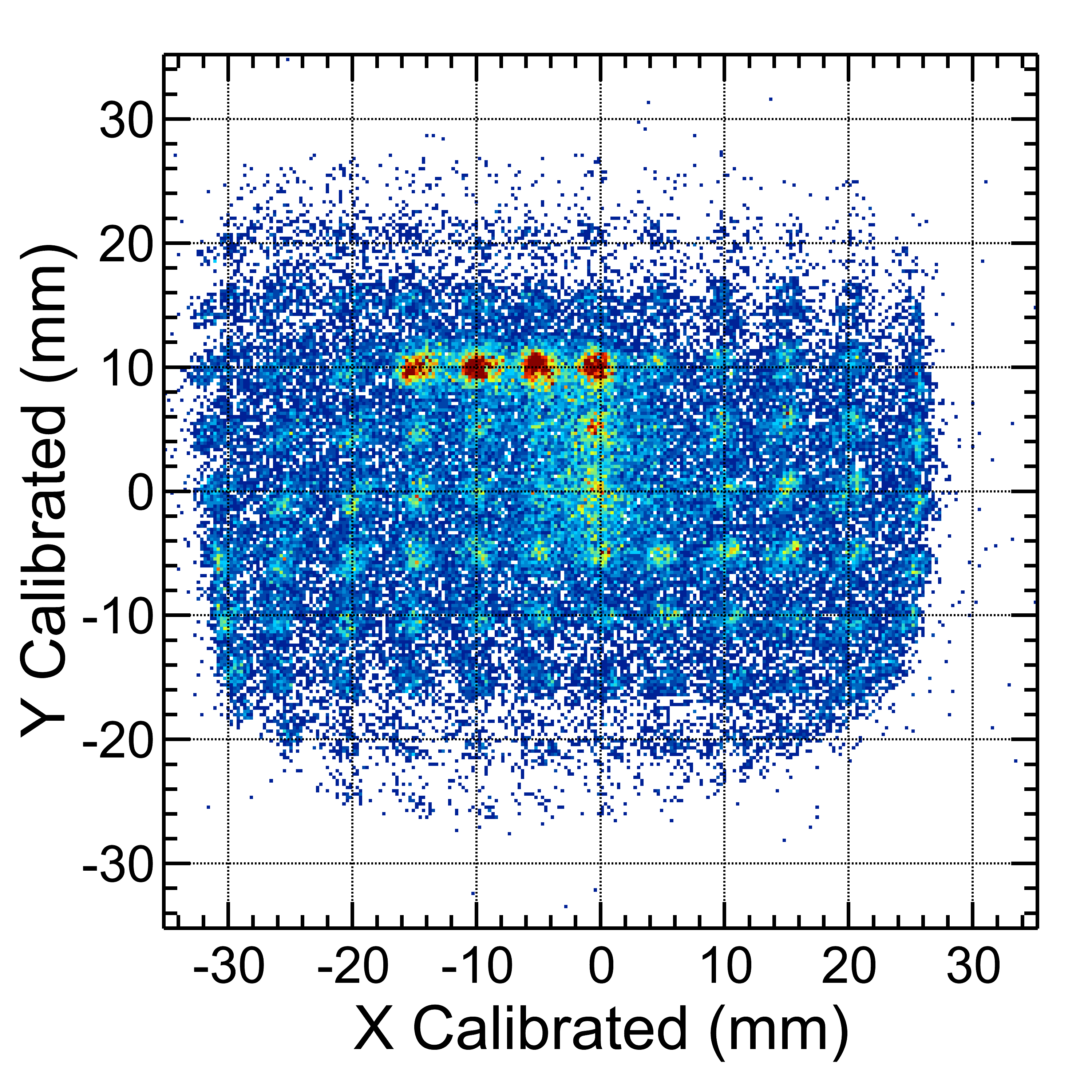}
   \phantomsubcaption\label{b}}
  \caption{(Color online). (Top) Raw and (bottom) calibrated position spectra using \num{72}-\si{MeV/nucleon} ${}^{70}$Se cocktail beam. The data was taken with magnetic field of $B_{\textrm{MCP}} = \SI{0.05}{T}$ and $B_{\textrm{foil}}=\SI{0.03}{T}$.}
  \label{fig:SeCalib}
\end{figure}

\autoref{fig:SeCalib} shows the uncalibrated and calibrated mask spectra from the ${}^{70}$Se secondary beam.
During the beam calibration test, the incoming secondary beam is defocused so that beam particles arrive at the target with nearly parallel trajectories and incoming angles close to \ang{0}.
The L-shaped pattern of the \num{1.5}-\si{mm} holes, as shown in \autoref{fig:AlphaCalib} for the $\alpha$-particle calibrations, can be clearly seen in \autoref{fig:SeCalib}.
Similar to the calibrated spectrum shown in \autoref{fig:AlphaCalib}, the calibration procedure correctly rotates the L-shaped pattern so that the center of the mask is located at $(0,0)$~\si{mm} and the spacing between holes is calibrated to the nominal \SI{5}{mm}.
The calibration procedure also corrects most of the non-linear distortions of the hole pattern at the edges of the uncalibrated spectrum.

There is a clear difference between the resolution of the position spectra for the $\alpha$ particles in \autoref{fig:AlphaCalib} and the ${}^{70}$Se secondary beam particles in \autoref{fig:SeCalib}.
The position resolution for the $\alpha$-source spectrum corresponding to the larger holes is approximately \num{1.5}-\si{mm} FWHM in both the vertical and horizontal directions.
In contrast, the position resolution for the data taken with the ${}^{70}$Se beam is approximately \num{2.5}-\si{mm} FWHM in both the vertical and horizontal directions.
After unfolding the influence of the hole size in the calibrated mask spectrum, the corresponding observed intrinsic resolution for the ${}^{70}$Se beam is about \num{2}-\si{mm} FWHM, approximately \num{2} times larger than the $\alpha$-source.

It is not obvious why the ${}^{70}$Se beam exhibits a resolution that is worse than that obtained with the $\alpha$ particles from the ${}^{228}$Th source.
The conditions for both measurements are similar.
The MCP foil voltages are set to \SI{-1000}{V}.
The beam counting rate is higher than the $\alpha$-particle rate, but there is no indication of a strong dependence of the resolution on the beam rate in the work of Shapira~et~al.~\cite{Shapira2000409,Shapira2000396}.

One possible difference is the particle energy.
The $\alpha$ particles range in energies from 1 to \SI{2}{MeV/nucleon} while the ${}^{70}$Se beam energy is \SI{72}{MeV/nucleon} before striking the MCP foils.
Kinetic energies of electrons scattered at angles greater \ang{60} from the MCP foil by $\alpha$ particles are estimated to be less than \SI{0.6}{keV}, while the corresponding kinetic energies of electrons scattered from the MCP foil by the ${}^{70}$Se beam can be up to \SI{37}{keV}.
Thus, a magnetic field that is adequate to provide good resolution with low-energy electrons may not be strong enough to confine the trajectories of the high-energy electrons to give similar position resolution.

In the following, we estimate this effect; the accuracy of our estimates, however, is limited by the fact that we do not have a measurement of the electron spectrum.
We therefore take the electron spectrum to be consistent with a primary distribution given by Mott scattering of electrons by the incident-beam particles~\cite{Mott:1965wt}.
Assuming the magnetic field varies slowly on the scale of the radius of the electron helical orbit, the motion can be assumed to be adiabatic and the final radius will be given by $r_{\textrm{MCP}} = r_{\textrm{foil}} \times \sqrt{B_{\textrm{foil}}/B_{\textrm{MCP}}}$, where $r_{\textrm{MCP}}$ and $B_{\textrm{MCP}}$ are the orbital radius and magnetic field at the MCP and $r_{\textrm{foil}}$ and $B_{\textrm{foil}}$ are the orbital radius and magnetic field at the foil~\cite{OttiniHustache:1999gg}.
The initial orbital radius $r_{\textrm{foil}}$ is determined by the component of the electron momentum perpendicular to the magnetic field $P_{\perp} = \sqrt{\frac{2}{3} (2m_{\textrm{e}}\langle E \rangle)}$, where $\langle E \rangle$ is the mean kinetic energy of emitted electron.

From the Mott scattering cross section, one can estimate the mean energy of primary electrons to be approximately $\langle E \rangle \approx E_{\textrm{min}} \ln(E_{\textrm{max}}/E_{\textrm{min}})$ where the relevant electron energies effectively range from an upper limit, $E_{\textrm{max}}$, given approximately by elastic scattering kinematics to the lower limit, $E_{\textrm{min}}$.
Our choice of \SI{0.040}{keV} for $E_{\textrm{min}}$ is approximate because the spectrum is dominated by secondary electrons.
It reflects a compromise between the strong production of secondary electrons in the target and the energy-range relationship for electrons~\cite{Jackson:2007ub,Ashley:1978tr}, which prevents the emission of most of them from the MCP foil (\num{290}-\si{\ug/cm^2} aluminized Mylar).
The energy for electrons back-scattered at $\theta > \ang{60}$ from the ${}^{70}$Se beam at \num{72}-\si{MeV/nucleon} incident energy can reach $E_{\textrm{max}}=\SI{37}{keV}$.
Due to dominance of the low-energy electrons, however, the mean scattered-electron energy is estimated to be approximately \SI{0.273}{keV} from the Mott scattering equation.
For the data in \autoref{fig:AlphaCalib} and \autoref{fig:SeCalib}, the magnetic field in the MCP assembly has values of $B_{\textrm{MCP}}=\SI{0.05}{T}$ and $B_{\textrm{foil}}=\SI{0.03}{T}$.
Thus, for collisions involving the ${}^{70}$Se beam, the FWHM contribution to the position resolution is about \SI{2.0}{mm}.

In the case when $\alpha$ particles of 1 to \SI{2}{MeV/nucleon} strike the MCP foil, we estimate that the primary electrons rarely exceed \SI{0.6}{keV} in energy and that most electrons are emitted from the surface of the foil.
Our estimate of the mean scattered-electron energy is \SI{0.1}{keV} and the FWHM contribution to the resolution is \SI{1.2}{mm}.
Based on the above calculations, the increase of scattered-electron energy with particle energy can degrade the position resolution of the microchannel-plate tracking detector.

\subsection{Position calibration with ${}^{56}$Ni cocktail beam at \SI{37}{MeV/nucleon}}
\label{sec:PosCal56Ni}
To further explore the influence of the primary electrons on the resolution, we examine the performance of these MCPs in another experiment involving a ${}^{56}$Ni beam at \num{37}-\si{MeV/nucleon} incident energy.
In the second experiment, both tracking detectors were equipped with stronger permanent magnets~\cite{company:msm};
The magnetic field measured at the MCP and the MCP foil positions are $B_{\text{MCP}}=\SI{0.21}{T}$ and $B_{\textrm{foil}}=\SI{0.13}{T}$ respectively, nearly quadruple the magnetic field of the preceding MCP setup.
Using a similar estimate as discussed in section 4.1, the mean scattered-electron energy from the ${}^{56}$Ni beam is approximately \SI{10}{keV}.
With the increased magnetic field, we estimate the FWHM contribution on the position resolution to be \SI{0.4}{mm}, better than the resolution obtained in the $\alpha$-source test.

\begin{figure}
  \centering
  {\includegraphics[width=0.40\textwidth]{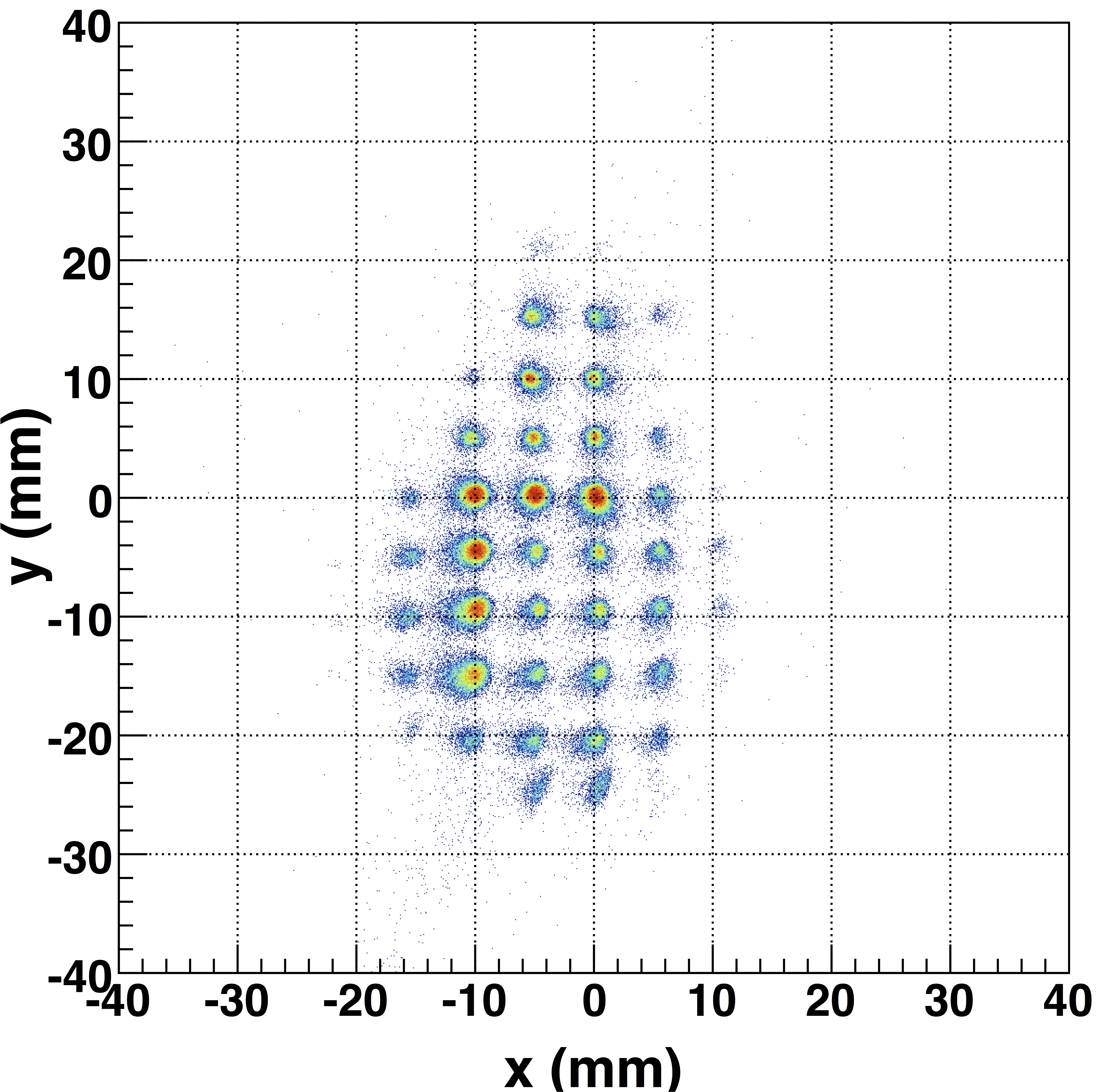}
   \phantomsubcaption\label{a}}
  {\includegraphics[width=0.40\textwidth]{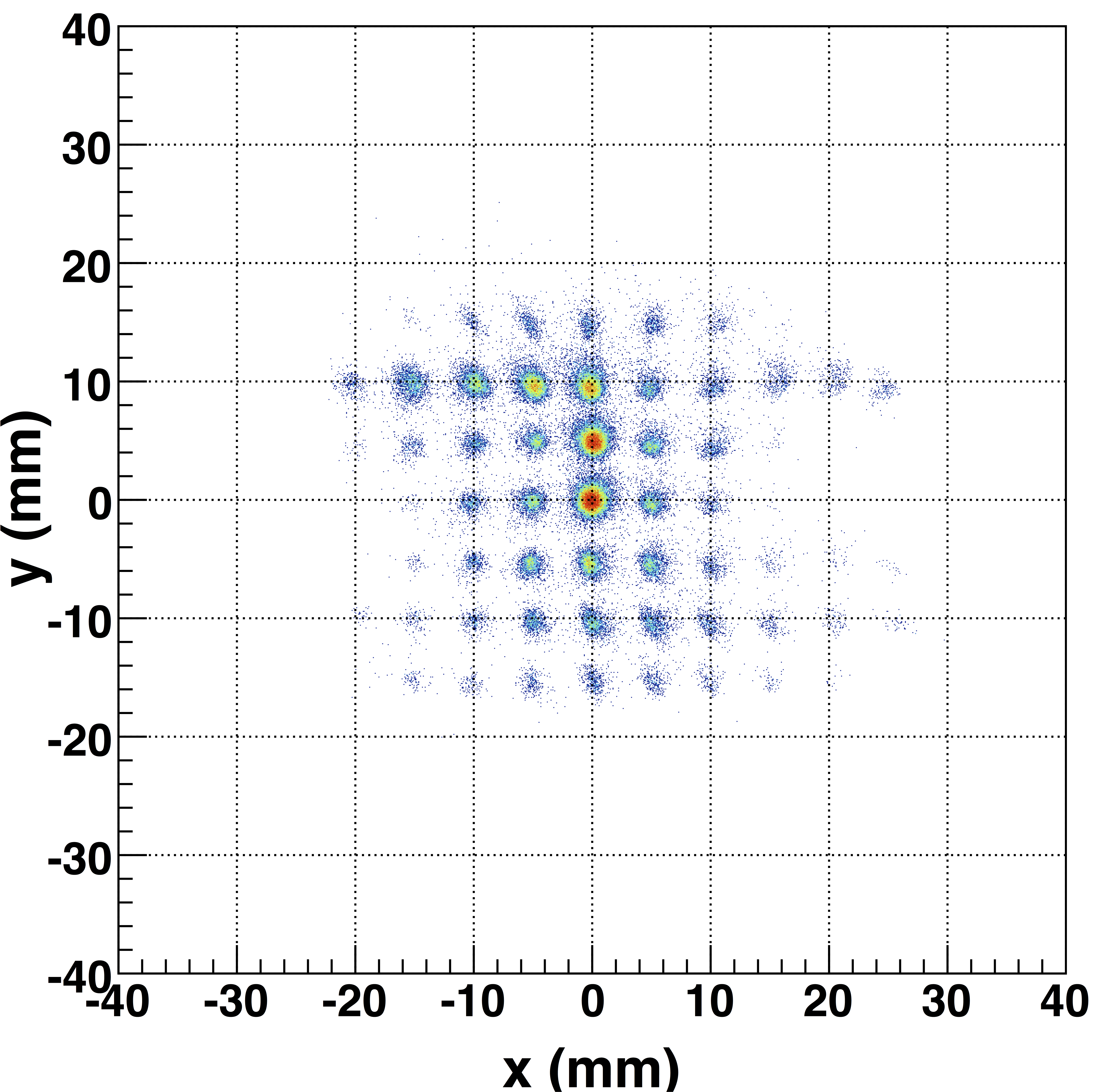}
   \phantomsubcaption\label{b}}
  \caption{(Color online). Calibrated position spectra for MCP0 (top) and MCP1 (bottom) obtained from the ${}^{56}$Ni cocktail beam. The data was taken with $B_{\textrm{MCP}}=\SI{0.21}{T}$ and $B_{\textrm{foil}}=\SI{0.13}{T}$.}
  \label{fig:NiCalib}
\end{figure}

\autoref{fig:NiCalib} displays the calibrated mask spectra for both MCPs with the ${}^{56}$Ni beam.
The resolution is significantly better than the observed resolution for the ${}^{70}$Se beam in \autoref{fig:SeCalib}.
Unfold the influence of the hole size in the mask for the ${}^{56}$Ni beam, a corresponding intrinsic resolution of \SI{1.1}{mm} is obtained.
The position resolution is similar to the resolution obtained with $\alpha$ particles from the ${}^{228}$Th source.
Thus, we see an improvement due to the use of a stronger magnetic field that is within our estimates given above.
The correlation, however, between predicted and measured trends is more suggestive than conclusive.
To be quantitative, one would need accurate measurements of the electron spectra emitted from the target for these three cases.

\section{Application to rare-isotope beam experiments}
\label{sec:AppRareIsoExp}
\begin{figure}
  \centering
  \includegraphics[width=0.450\textwidth]{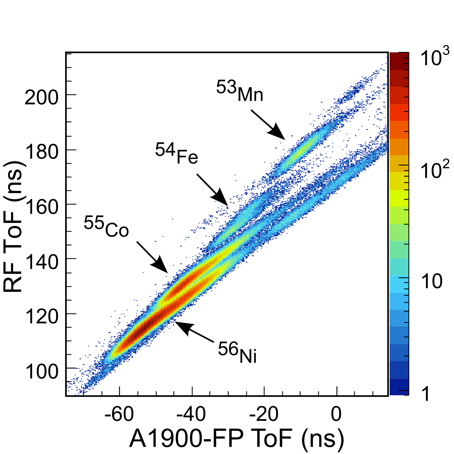}
  \caption{(Color online). Secondary-beam particle identification based on time-of-flight signals from the cyclotron RF and a A1900 scintillator.}
  \label{fig:beamID}
\end{figure}

In this section, we discuss the position resolution we obtained in an experiment in which we measured the differential cross sections for the $p$($^{56}$Ni,$d$)$^{55}$Ni transfer reactions~\cite{Sanetullaev:vv}.
A mixed secondary beam consisting of \SI{70}{\percent} ${}^{56}$Ni at \SI{37}{MeV/nucleon} was produced through projectile fragmentation of a ${}^{58}$Ni primary beam at NSCL.
Other major isotopes in the cocktail beam are nuclei containing the same number of neutrons such as ${}^{55}$Co and ${}^{54}$Fe, as shown in \autoref{fig:beamID}.
These beams had similar magnetic rigidities to ${}^{56}$Ni, thus they pass through the A1900 fragment separator to the reaction chamber.
The isotopes composing the secondary beam are identified event-by-event by comparing their measured time-of-flight (ToF) signals given by the cyclotron RF and a scintillator at the A1900 extended focal plane (XFP) relative to the timing signal generated by the S800.
A gate on the ${}^{56}$Ni beam has been imposed on all the data discussed in this section.

The transfer reaction $p$(${}^{56}$Ni,$d$)${}^{55}$Ni is produced by bombarding a polyethylene (CH$_{2}$)$n$ target with \num{37}-\si{MeV/nucleon} ${}^{56}$Ni beam.
Deuterons produced in the reactions are detected by the High Resolution Array, HiRA~\cite{2007NIMPA.583..302W}, and the associated heavy ${}^{55}$Ni residue is detected in coincidence using the S800 spectrograph~\cite{bazin2003ss}.
In order to resolve the energy states of the ${}^{55}$Ni residuals at high resolution, accurate determinations of the deuteron energy and emission angle are required.
The \num{16} HiRA telescopes were located \SI{35}{cm} downstream of the target.
They are arranged to subtend polar laboratory angles of \ang{5}--\ang{40}.
Each telescope consists of a single-sided $\Delta$E (\SI{65}{\um}) and a double-sided E (\SI{1500}{\um}) silicon strip detector, with active areas of $\SI{6.25}{cm}\times\SI{6.25}{cm}$, backed by four separate CsI(Tl) crystals mounted in quadrants.
The $32\times32=\SI{1024}{pixels}$ defined by the vertical front and horizontal back strips in the double-sided silicon E detector allowed the angles of emitted deuterons to be determined, relative to the center of the target, with a precision of \ang{\pm0.16}.

The actual laboratory emission angle for the deuterons, however, depends on the position where the reaction occurs in the target and on the momentum direction of the incident ${}^{56}$Ni beam before the reaction.
The MCP tracking system is used to constrain these quantities.
This system employed two microchannel-plate tracking detectors, MCP0 and MCP1, which measure the position of the beam at two points along the incoming beam trajectory.
The detectors are placed \SI{50}{cm} apart, with MCP0 and MCP1 denoting the upstream and downstream MCPs, respectively.
The reaction target is located \SI{10}{cm} downstream of MCP1 and is perpendicular to the beam axis.
Each tracking-detector assembly is rotated by \ang{60} with respect to the beam so as not to obstruct the incoming beam particles.
Due to the sub-nanosecond timing resolution of the MCPs, MCP1 also provides a time signal for residues detected in the focal plane detectors of the S800 mass spectrometer.
The time-of-flight and energy loss of the residues in the S800 provides good identification of ${}^{55}$Ni particles, which is critical in the analysis.
Additional details on the experimental set up can be found in Ref.~\cite{2011PhRvL.106y2503R}.

The strong magnetic fields in the MCP setup necessary for increasing position resolution, however, influence the trajectories of the ${}^{56}$Ni beam ions.
For example, the first MCP magnet deflects the beam by an angle of $\theta=B d \sin(\gamma)/B\rho$.
Here, $B$ is the magnetic field at the MCP foil ($B_{\textrm{foil}}=\SI{0.13}{T}$), $d$ is the distance the beam traveled in the magnetic field \SI[separate-uncertainty]{8\pm0.5}{cm}), $\gamma$ is the angle of the foil with respect to the normal (\ang{60}), and $B\rho$ is the magnetic rigidity of the beam (\SI{1.72}{Tesla\cdot{}m}).
The estimated deflection angle is \ang{0.17}.
To compensate for this deflection, the magnetic fields of the two MCPs were arranged to have opposite polarity with respect to each other.
In this configuration, the main overall effect of the two sets of magnets on the beam is an upward shift by \SI{1.5}{mm}.
Because the experiment only depends on the position and angle of the beam at the target, this shift is of no consequence as long as it is small compared to the beam spot.

\subsection{Reaction target beam-tracking reconstruction}
\label{sec:ReacTarRecon}
\begin{figure}
  \centering
  \includegraphics[width=0.40\textwidth]{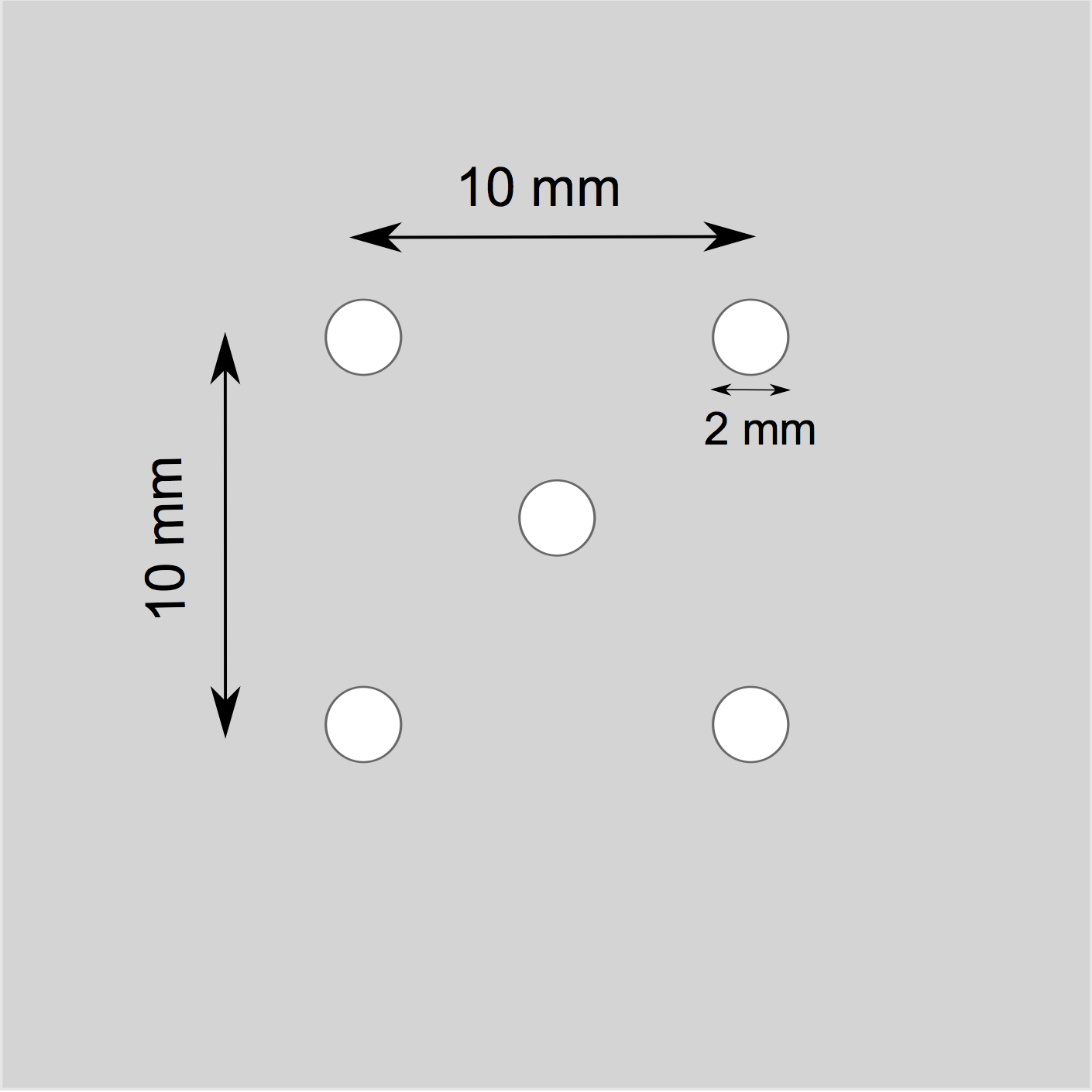}
  \caption{A schematic view of the reaction-target mask.}
  \label{fig:targetMask}
\end{figure}

To evaluate the uncertainties in the beam-tracking reconstruction and the MCP calibration procedure, a mask was inserted at the reaction-target position.
The mask is made out of a \num{1.6}-\si{mm} thick aluminum plate with five \num{2}-\si{mm} holes, as shown in the schematic diagram of \autoref{fig:targetMask}.
The four outer holes form a square with \num{10}-\si{mm} sides while a fifth hole is located at the mask center.
Again, as with the MCP beam calibrations, the S800 spectrograph is used to detect coincident particles that pass through the holes in the mask.

\begin{figure}
  \centering
  \includegraphics[width=0.450\textwidth]{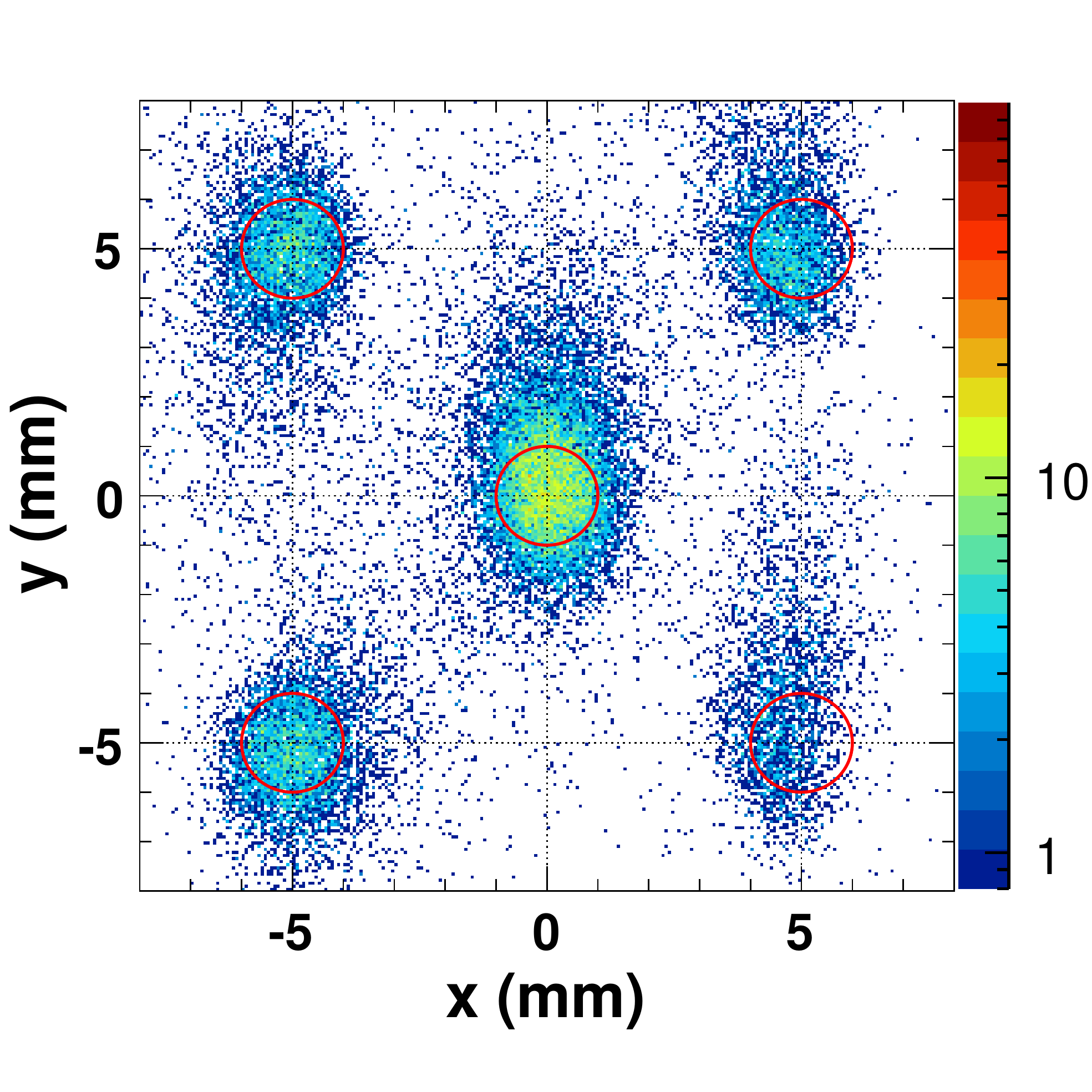}
  \caption{Calculated position at the reaction target using measured MCP tracking data. The overlaid circles represent the reaction-target mask holes from \autoref{fig:targetMask}.}
  \label{fig:MCPNiTargetMask}
\end{figure}

From the position data of MCP0 and MCP1 the beam is tracked, and the individual beam-particle trajectories extrapolated to the reaction-target location.
A comparison of the data obtained with the mask inserted to the actual mask holes is shown in \autoref{fig:MCPNiTargetMask}.
The calculated positions reproduce the reaction-target mask reasonably well, but the resolution of reconstructed positions at the target is worse than the position resolution for the individual MCP mask calibrations.
For the outer holes, the resolutions are \SI{1.2}{mm} in the horizontal direction and \SI{1.5}{mm} in the vertical direction.
The difference in the horizontal and vertical resolutions is mainly due to the rotation of the MCP foils.
The distribution of the errors is not uniform, however, and is somewhat worse at the center of the target ($\sim\SI{2}{mm}$) than at the edges.
Also, the resolution of the left side is better than the right side. We note that the average resolution is comparable to the \num{2}-\si{mm} pitch of the HiRA silicon-strip detectors.

\begin{figure}
  \centering
  \includegraphics[width=0.450\textwidth]{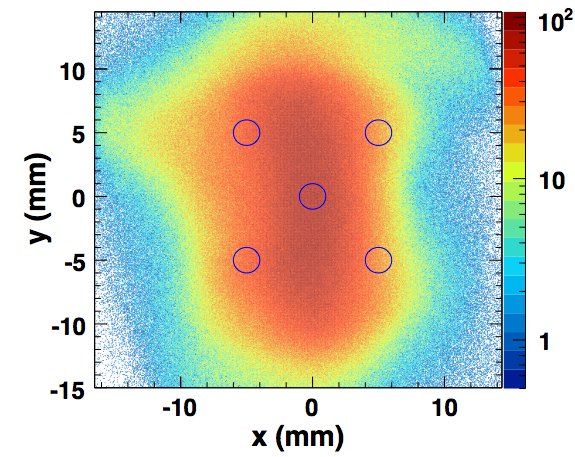}
  \caption{Position spectrum of the ${}^{56}$Ni beam on the reaction target reconstructed by the MCP tracking system. The circles show the reaction-target mask holes from \autoref{fig:targetMask}.}
  \label{fig:NiBeam}
\end{figure}

\autoref{fig:NiBeam} shows a position spectrum of the ${}^{56}$Ni beam at the reaction target, extrapolated from the tracking positions measured at the two MCPs.
For this experiment the ${}^{56}$Ni secondary beam has dimensions of \num{11}-\si{mm} FWHM in the horizontal direction and \num{17}-\si{mm} FWHM in the vertical direction.
The target mask from \autoref{fig:targetMask} is also superimposed on the spectrum.
Comparing \autoref{fig:MCPNiTargetMask} and \autoref{fig:NiBeam}, one can estimate that the tracking system yields an improvement in the interaction position by a factor of $\sim10$ better than the width of the beam spot.
If we made no correction for the interaction point on the target, particles emitted from reactions occurring at the target center and at a position $1\sigmaÊ=Ê\SI{7.3}{mm}$ away from the beam spot center would have significantly different scattering angles of approximately \ang{1.2}, even if they are detected in the same pixel of the HiRA device.

\begin{figure}
  \centering
  \includegraphics[width=0.450\textwidth]{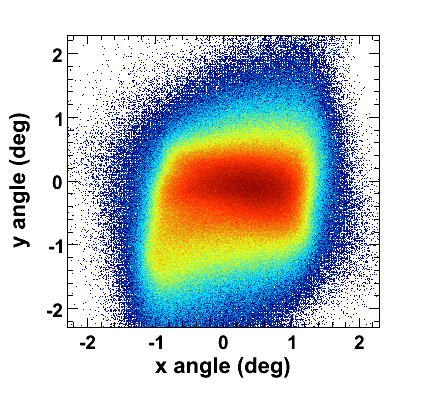}
  \caption{Beam angle spectrum of the ${}^{56}$Ni beam on the reaction target reconstructed by the MCP tracking system.}
  \label{fig:NiBeamAngle}
\end{figure}

The scattering angle of an emitted particle after a nuclear reaction also depends on the incident-beam angles at the reaction target.
\autoref{fig:NiBeamAngle} shows the spectrum of the incident-beam angles.
The angular distributions have widths of $\delta\theta_{x}^{\textrm{beam}}=\ang{1.5}$~FWHM and $\delta\theta_{y}^{\textrm{beam}}=\ang{1.35}$~FWHM.
Folding the beam angle distribution against the target spot-size distribution, one has $\delta\theta_{x}^{\textrm{total}}=\ang{2.3}$~FWHM and $\delta\theta_{y}^{\textrm{total}}=\ang{3.0}$~FWHM, respectively.
Most transfer-reaction experiments are designed to achieve an angular resolution of less than \ang{0.3}, thus the target beam-spot size and beam angular resolution are critical issues if uncorrected.

\subsection{Improvements in angular and energy resolution}
\label{sec:Improve}
\begin{figure}[t]
  \centering
  {\includegraphics[width=0.40\textwidth]{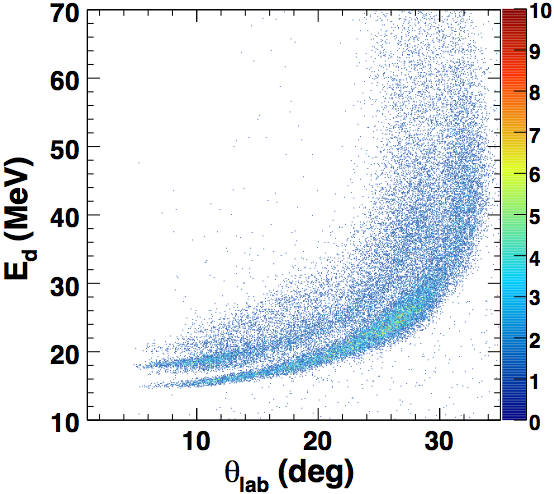}
   \phantomsubcaption\label{a}}
  {\includegraphics[width=0.40\textwidth]{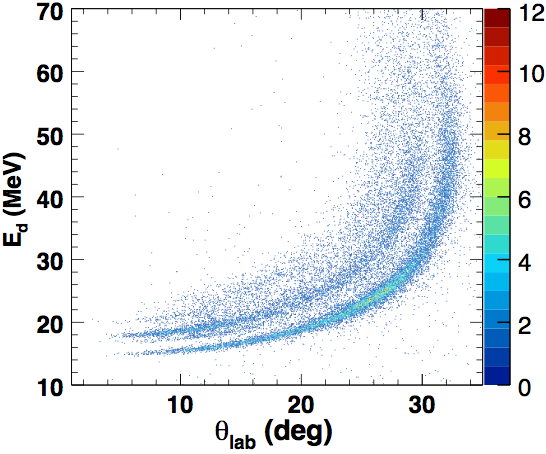}
   \phantomsubcaption\label{b}}
  \caption{(Color online). Deuteron energy as a function of $\theta_{\textrm{lab}}$ without MCP correction (top) and with MCP correction (bottom).}
  \label{fig:NiKin}
\end{figure}

In the neutron-transfer experiment energy and angles of the deuteron are measured along with the momentum of the residue,${}^{55}$Ni, from the $p$(${}^{56}$Ni,$d$)${}^{55}$Ni reaction. 
Since the ${}^{56}$Ni beam energy is fixed and the target is at rest, conservation of energy and momentum dictates that the kinematic relationship between $E_d$ and $\theta_d$ forms a hyperbolic curve for a fixed energy state of ${}^{55}$Ni.
\autoref{fig:NiKin} shows a two-dimensional spectrum of the measured lab angle versus lab energy of the emitted deuterons. 
  
Using beam position and angle information from the MCP tracking system, the deuteron energy and scattering angle is corrected as shown in the lower spectrum in \autoref{fig:NiKin}.
Both the ground state and the excited state at \SI{3.185}{MeV}~\cite{Junde:2008dl} are clearly observed.
At angles less than \ang{15}, the first excited-state peak at \SI{2.089}{MeV} is identified in the bottom panel of \autoref{fig:NiKin} after the data has been corrected using the MCP tracking.
The improvement in energy resolution is minimal ($\sim\SI{20}{\percent}$ from \SI{0.5}{MeV} to {0.4}{MeV}) at forward angles, but increases by a factor of two (from \SI{3}{MeV} to \SI{1.5}{MeV}) at backward angles. Improvement of the resolution obtained in the center-of-mass will be described in detail in a forth coming paper discussing the physics of the experiment.

The current energy resolution is limited by the target thickness of \SI{9.6}{mg/cm^2}.
A thick target is needed to obtain sufficient counts for the first excited state.
If a thinner target such as \SI{2.5}{mg/cm^2} is used, the ground-state energy resolution would be better than \SI{0.4}{MeV} at the forward angles.
In future studies of states at higher excitation energy, thinner targets could be used to maximize resolution.

\section{Summary}
\label{sec:Summary}
We have successfully used devices based on microchannel plates with position-sensitive resistive anodes to track fast heavy-ion beam particles.
Strong magnetic fields on the order of \SI{2}{kG} were required, however, to obtain position resolution of $\sim\SI{1}{mm}$ for fast ($\SI{>35}{MeV/nucleon}$) heavy ions such as ${}^{56}$Ni.
Our experience from a $p$(${}^{56}$Ni,$d$)${}^{55}$Ni transfer experiment suggests the MCPs can withstand incident-beam intensities of $\sim\SI{5e5}{pps}$ for up to \SI{10}{days} before degradation in resolution or the microchannel plates occurs.
Such features make MCPs desirable as tracking detectors for experiments requiring high intensity rare-isotope beams.

%

\section{Acknowledgements}
\label{sec:Ack}
We wish to acknowledge the support of the National Science Foundation Grants No. PHY-0606007, PHY-0855013, and PHY-1064280.





\bibliographystyle{model1-num-names}
\bibliography{MCPbibliographyStandard}











\end{document}